\documentclass[cernpreprint,atlasdraft=false,texlive=2016,UKenglish,texmf]{atlasdoc}
 
\usepackage{atlaspackage}
\usepackage[backref=false,maxbibnames=3]{atlasbiblatex}
 
\usepackage{wrapfig}
\usepackage{atlasphysics}
 
\graphicspath{{logos/}{figures/}}
 
\usepackage{ANA-TOPQ-2018-29-PAPER-defs}
\usepackage{PMGRefs-defs}

\PreprintIdNumber{CERN-EP-2020-139}

\newcommand{\AtlasCoordFootnote}{
ATLAS uses a right-handed coordinate system with its origin at the nominal interaction point (IP)
in the centre of the detector and the \(z\)-axis along the beam pipe.
The \(x\)-axis points from the IP to the centre of the LHC ring,
and the \(y\)-axis points upwards.
Cylindrical coordinates \((r,\phi)\) are used in the transverse plane,
\(\phi\) being the azimuthal angle around the \(z\)-axis.
The pseudorapidity is defined in terms of the polar angle \(\theta\) as \(\eta = -\ln \tan(\theta/2)\).
Angular distance is measured in units of \(\Delta R \equiv \sqrt{(\Delta\eta)^{2} + (\Delta\phi)^{2}}\).}
 
\AtlasTitle{Test of the universality of $\tau$~and $\mu$~lepton couplings in $W$-boson decays with the ATLAS detector}
 
\AtlasVersion{4.2}
\AtlasJournal{Nature Physics}
\AtlasJournalRef{Nat. Phys. 17 (2021) 813}
\AtlasJournalRef{Nat. Phys. 17 (2021) 813}
\AtlasDOI{10.1038/s41567-021-01236-w}
\AtlasAbstract{
The Standard Model of particle physics encapsulates our best current
understanding of physics at the smallest scales. A fundamental axiom
of this theory is the universality of the couplings of the different
generations of leptons to the electroweak gauge bosons.  The
measurement of the ratio of the decay rate of $W$ bosons to
$\tau$-leptons and muons, $R(\tau/\mu)$, constitutes an important test
of this axiom.
Using 139~fb$^{-1}$~of proton--proton collisions recorded with the ATLAS
detector at a centre-of-mass energy of 13~TeV, we report a
measurement of this quantity from di-leptonic $t\bar{t}$ events where
the top quarks decay into a $W$ boson and a bottom quark.  We can
distinguish muons originating from $W$ bosons and those originating
from an intermediate $\tau$-lepton
through the muon transverse impact parameter and differences in the muon transverse momentum spectra.
The measured value of
$R(\tau/\mu)$~is $0.992 \pm 0.013~[\pm
0.007~\textrm{(stat)}~\pm~0.011~\textrm{(syst)}]$~and is in agreement
with the hypothesis of universal lepton couplings as postulated in the
Standard Model. This is the only such measurement from the Large Hadron Collider, so far, and obtains twice the precision of previous measurements.}
 
\author{The ATLAS Collaboration}
 
\AtlasRefCode{TOPQ-2018-29}

\hypersetup{pdftitle={ATLAS document},pdfauthor={The ATLAS Collaboration}}
 
\begin{document}
 
\maketitle

\renewcommand{\textfraction}{0.001}

% The next lines are included from the .//sections/IntroDetData.tex input file
\section*{Main}

It is a fundamental axiom of the Standard Model
(SM) that the couplings of the electroweak gauge bosons ($W,Z$) to
charged leptons, $g_\ell$ ($\ell = $~electron ($e$), muon ($\mu$), tau ($\tau$)), are independent of
the mass of the leptons. This fundamental assumption is referred to as
\textit{lepton-flavour universality} and is tested in this paper by
comparing the relative decay rates, or branching ratios ($B$), of on-shell W-bosons to $\tau$-leptons and muons, by measuring the ratio $\Rtmu = B(W\rightarrow\tau\nu_\tau)/B(W\rightarrow \mu\nu_\mu)$.

The measurement exploits the large number of top and anti-top quark
pair (\ttbar) events produced in proton-proton ($pp$) collisions at
the Large Hadron Collider (LHC).  Given that $B(t\rightarrow
Wq)$ is close to 100\%, this gives
a very large sample of $W$ boson pairs. These are used to obtain a large sample of
clean and unbiased $W$ boson decays to muons and $\tau$-leptons.  The
$\tau$-leptons are identified through their decay to muons. The $\tau$-lepton has a significant lifetime~\cite{PhysRevLett.48.66,Belous:2013dba} compared to the $W$ boson and undergoes a multi-body decay to a muon and neutrinos. This leaves two distinctive signatures in the detector, the displacement of the $\tau$~decay vertex and on average a lower muon
transverse momentum (\pT). These features are used to distinguish between
muons from the $W\rightarrow\tau\nu_\tau\rightarrow
\mu\nu_\mu\nu_\tau\nu_\tau$ and $W\rightarrow \mu\nu_\mu$ processes,
to extract \Rtmu. The precision of the measurement relies on the highly accurate
reconstruction of muon tracks obtainable by the ATLAS experiment.
 
Previously, \Rtmu has been measured at the
Large Electron--Positron Collider (LEP), yielding a combined value of
$1.070\pm0.026$~\cite{Schael:2013ita}. The SM expectation of \Rtmu~is unity (neglecting very small phase space effects due to the
masses of the decay products ($\sim5\times10^{-4}$)~\cite{Tanabashi:2018oca}) which differs from the measured value, motivating a
precise measurement of this ratio at the LHC. Other experimental
measurements of the ratio $B(W\rightarrow \tau\nu_\tau)/B(W\rightarrow
\ell\nu_\ell)$, where $\ell$~is either an electron or a muon, have not
yet reached the precision of the LEP results
\cite{Albajar:1988ka,Abe:1991fb,Alitti:1992hv,Abbott:1999pk,Sirunyan:2019guq}. The
equivalent ratio for the two light generations, $B(W\rightarrow
\mu\nu_\mu)/B(W\rightarrow e\nu_e)$, has been accurately measured by
the LEP \cite{Schael:2013ita}, LHCb \cite{Aaij:2016qqz} and ATLAS
\cite{STDM-2012-20} experiments, and is found to be consistent with
the SM prediction at the 1\% level. Additionally, while most
low-energy experiments \cite{Pich:2013lsa} show good agreement, to very high precision,
with the hypothesis of universality of lepton couplings, recent results from LHCb
\cite{Aaij:2017uff,Aaij:2017deq,2020PhRvL.125a1802A,2016JHEP...02..104A},
Belle \cite{Hirose:2016wfn,Hirose:2017dxl,Abdesselam:2019dgh} and
BaBar \cite{Lees:2012xj,Lees:2013uzd} show some tension with the SM,
further motivating this analysis.
 
This measurement relies on precise knowledge of the branching ratio of
$\tau$-leptons decaying to muons to extrapolate to the full
$W\rightarrow\tau\nu_\tau$~branching ratio. The value of
$(17.39\pm0.04)\%$ measured by the LEP experiments
\cite{Schael:2005am,Abbiendi:2002jw,Acciarri:2001sg,Abreu:1999rb,Tanabashi:2018oca}
is used in the analysis.

\section*{Experimental set-up}\label{sec:detector}
The ATLAS experiment~\cite{PERF-2007-01,Capeans:1291633,PIX-2018-001}
at the LHC is a multipurpose particle detector with nearly hermetic
coverage for recording particles produced in LHC collisions through a
combination of particle position and energy measurements.  ATLAS has a
forward--backward symmetric cylindrical geometry and a near \(4\pi\)
coverage in solid angle. More details of the ATLAS coordinate system
can be found in \textit{Methods}.  ATLAS consists of an inner tracking
detector surrounded by a thin superconducting solenoid providing a
\SI{2}{\tesla} axial magnetic field, electromagnetic and hadron
calorimeters, and a muon spectrometer.  The inner tracking detector
covers the pseudorapidity range \(|\eta| < 2.5\).  It consists of
silicon pixel, silicon microstrip, and transition radiation tracking
detectors, and the innermost layer of the pixel detector is at a
radius of 33 mm from the beamline providing precise measurement of
track impact parameters.  Lead/liquid-argon (LAr) sampling
calorimeters provide electromagnetic (EM) energy measurements with
high granularity.  A steel/scintillator-tile hadron calorimeter covers
the central pseudorapidity range (\(|\eta| < 1.7\)).  The endcap and
forward regions of the detector are instrumented with LAr calorimeters
for both EM and hadronic energy measurements up to \(|\eta| = 4.9\).
The muon spectrometer surrounds the calorimeters and is based on three
large air-core toroidal superconducting magnets with eight coils each.
The field integral of the toroids ranges between \num{2.0} and
\SI{6.0}{\tesla\metre} across most of the detector.  The muon
spectrometer includes a system of precision tracking chambers and fast
detectors for triggering. A two-level trigger system is used to select
events.  The first-level trigger is implemented in hardware and uses a
subset of the detector information to accept events at a rate of
\SI{100}{\kHz}.  This is followed by a software-based trigger that
reduces the accepted event rate to \SI{1}{\kHz} on average, depending
on the data-taking conditions.
 
The analysed $pp$ collision data were recorded with the ATLAS detector
from 2015 to 2018 at a centre-of-mass energy of $\sqrt{s}=13$~TeV. Due
to the conditions required to achieve high instantaneous luminosity,
in these data there are additional collisions in the same and
neighbouring LHC proton bunch crossings (pile-up). This resulted in an
average of 34 collisions per bunch crossing.
 
Events in this measurment were selected by single-lepton
triggers \cite{TRIG-2016-01,TRIG-2018-05,Aad:2020uyd} requiring a
single high-\pT isolated electron or muon. After the application of
data-quality requirements~\cite{DAPR-2018-01}, the data sample
corresponds to an integrated luminosity of 139~\ifb~with an
uncertainty of 1.7\%~\cite{ATLAS-CONF-2019-021}~obtained using the
LUCID-2 detector \cite{LUCID2} for the primary luminosity
measurements.
 
Samples of simulated events were produced using Monte Carlo (MC) techniques to model
the different Standard Model processes. After event generation of the
process of interest for each sample, the detector
response was modelled using a simulation based on
\textsc{Geant4}~\cite{Agostinelli:2002hh}. The data and MC simulated
events were passed through the same reconstruction and analysis
procedures. Samples were simulated for the signal processes, the
production of $\ttbar$ and single top quarks in association with a $W$
boson ($Wt$), as well as the different backgrounds.
Precise details of the theoretical predictions and event generators used to create the simulated events is given in
\textit{Methods}.
% End of text imported from the .//sections/IntroDetData.tex input file
% The next lines are included from the .//sections/SelectCalibNorms.tex input file
 
\section*{Selecting a pure and unbiased sample of $W$ bosons}
\label{sec:selection}
 
Collisions are selected in order to obtain a high purity sample of
\ttbar events in which the top quarks decay into a $W$ boson and a
$b$-quark, and the two $W$ bosons then decay to leptons. This is referred to
as di-leptonic \ttbar. The two leptonic $W$ boson decays are exploited
in a \textit{tag and probe} approach: in each event \textit{tag}
leptons (electrons or muons) are used to select the event, after
which a test is performed to determine, in an unbiased way, whether the probe muon was directly produced by a $W$~boson, $W\rightarrow\mu\nu_\mu$,
or via an intermediate $\tau$-lepton,
$W\rightarrow\tau\nu_\tau\rightarrow\mu\nu_\mu\nu_\tau\nu_\tau$. Events
are categorised into \textit{signal} regions, used to extract \Rtmu,
and additional \textit{control} regions, used to constrain the
normalisation of the major backgrounds.
 
The selections rely on reconstructed muons,
electrons and hadronic jets.
Details of the physics object reconstruction defintions used are given in
\textit{Methods}. Events entering the signal region are required to
contain either one electron and one muon of opposite electric charge
($e$--$\mu$ channel), or two muons of opposite electric charge
($\mu$--$\mu$ channel). In the $e$--$\mu$~channel the electron is required to pass the trigger and in the
$\mu$--$\mu$~channel the \textit{tag} muon is required to pass the trigger.
This ensures that the \textit{probe} muons have no
trigger bias. If both leptons in the $\mu$--$\mu$~channel satisfy the
\textit{tag} and \textit{probe} criteria, both muons in turn are used as
\textit{probes}. Events with more than two leptons are rejected. In
addition, events must have at least two reconstructed hadronic jets
which are identified as containing a $b$-hadron. Finally, to reduce
backgrounds from $Z$ bosons and hadron decays, events where the invariant mass of
the two muons, $m_{\mu\mu}$, satisfies $85<m_{\mu\mu}<95$~GeV are excluded in the $\mu$--$\mu$~channel,
and events with di-lepton mass $m_{\ell\ell}<15$~GeV are excluded in
both channels.
 
This selection results in a sample of approximately half a million
collision events, with a di-leptonic \ttbar purity of over 95\% (85\%) in the $e$--$\mu$ ($\mu$--$\mu$) channel. This
sample is used to test the origin of the \textit{probe} muons and
extract the measurement of \Rtmu. This is the ratio of the
number of events in which the \textit{probe} muon orginates from the
process $W\rightarrow\tau\nu_\tau\rightarrow
\mu\nu_\mu\nu_\tau\nu_\tau$, referred to as \taumu, to those which
come from the process $W\rightarrow \mu\nu_\mu$, referred to as
\promptmu. A fit is performed 
which exploits the difference in shape between these two components
and the backgrounds of the distributions of the \textit{probe} muon's
transverse impact parameter $|d_0^\mu|$ and its transverse momentum
$\pT^\mu$.

\section*{Muon transverse impact parameter distribution calibration}
\label{sec:calib}
 
The muon's transverse impact parameter $|d_0^\mu|$ has particular
importance for this analysis and requires careful treatment. It is
measured in the $x$--$y$~plane as the closest distance of approach of the
track to the beamline. The shape of the $|d_0^\mu|$~distribution of prompt
muons is determined using a $Z\rightarrow\mu\mu$~\textit{calibration}
region to create templates that are used to predict the distribution in the signal region.
 
The calibration region is defined by requiring two muons
satisfying the same kinematic criteria as in the signal region, but with
the di-muon mass requirement changed to $85<m_{\mu\mu}<100$~GeV. No
requirements on hadronic jets are applied. This gives a sample of
approximately 95 million prompt muons with a purity of $>99.9\%$.
 
Templates of the shape of the $|d_0^\mu|$~distribution are then taken from this data
sample after subtracting the expected contributions from the
simulation of processes with significant muon parent lifetimes, primarily
$Z\rightarrow\tau\tau$. These $|d_0^\mu|$ templates are
extracted in 33 bins in $\pT^\mu$~and $|\eta^{\mu}|$~to capture the
dependence of the distribution on these variables. Separate templates
are used for 2015+2016, 2017 and 2018 data to account for differences
in the beam conditions and in the alignment of the inner detector.
 
Additionally, using this calibration region, the Gaussian part of the
$|d_0^\mu|$~resolution is estimated in data and simulation by fitting
the $|d_0^\mu|$~distribution in the range $|d_0^\mu|<0.02$~mm. For
$\pT^\mu = 20$~GeV the resolution is approximately 14
$\mu$m. Corrections to account for differences in the resolution of
the detector between the data and simulation are applied to the
muons from $\tau$~decays and hadron decays. For the
range of $|d_0^\mu|$~values considered in this analysis, $|d_0^\mu|<0.5$~mm, the
resolution measured from prompt muons is applicable to those with
significant displacement.

\section*{Background normalisation estimation}
\label{sec:background}
 
The two largest backgrounds are $Z(\rightarrow\mu\mu)$+jets and events
in which the \textit{probe} muon does not originate from a $W$ boson
decay. Three dedicated control regions are used to extract the
normalisation of these backgrounds.

The $Z(\rightarrow\mu\mu)$+jets background is important at small
values of $|d_0^\mu|$. The normalisation of the
$Z(\rightarrow\mu\mu)$+jets background in the $\mu$--$\mu$~channel is
extracted from the data in a control region where the same event
selection is applied, including the hadronic jet requirements, but
without the $m_{\mu\mu}$ criterion and is then extrapolated to the
signal region using simulation. In the control region the peak of the
invariant mass distribution of the dimuon system is fitted over the
range $50<m_{\mu\mu}<140$~GeV. A Voigt profile \cite{voigt1912uber} is
used for the $Z\rightarrow\mu\mu$~resonance and a third-order
Chebychev polynomial for all non-resonant processes, which provides a good
description of the data. Other functions were tested to provide a
systematic uncertainty which is combined with the statistical
uncertainties. The normalisation factor required to scale the
simulated sample to data is found to be $1.36\pm0.01$.
The di-muon invariant mass in the control region is shown in Figure \ref{fig:control.a} after this
normalisation is applied. This normalisation factor is also applied to
the small $Z(\rightarrow\tau\tau$)+jets background.
 
\begin{figure}
\centering \includegraphics[width=5.1cm]{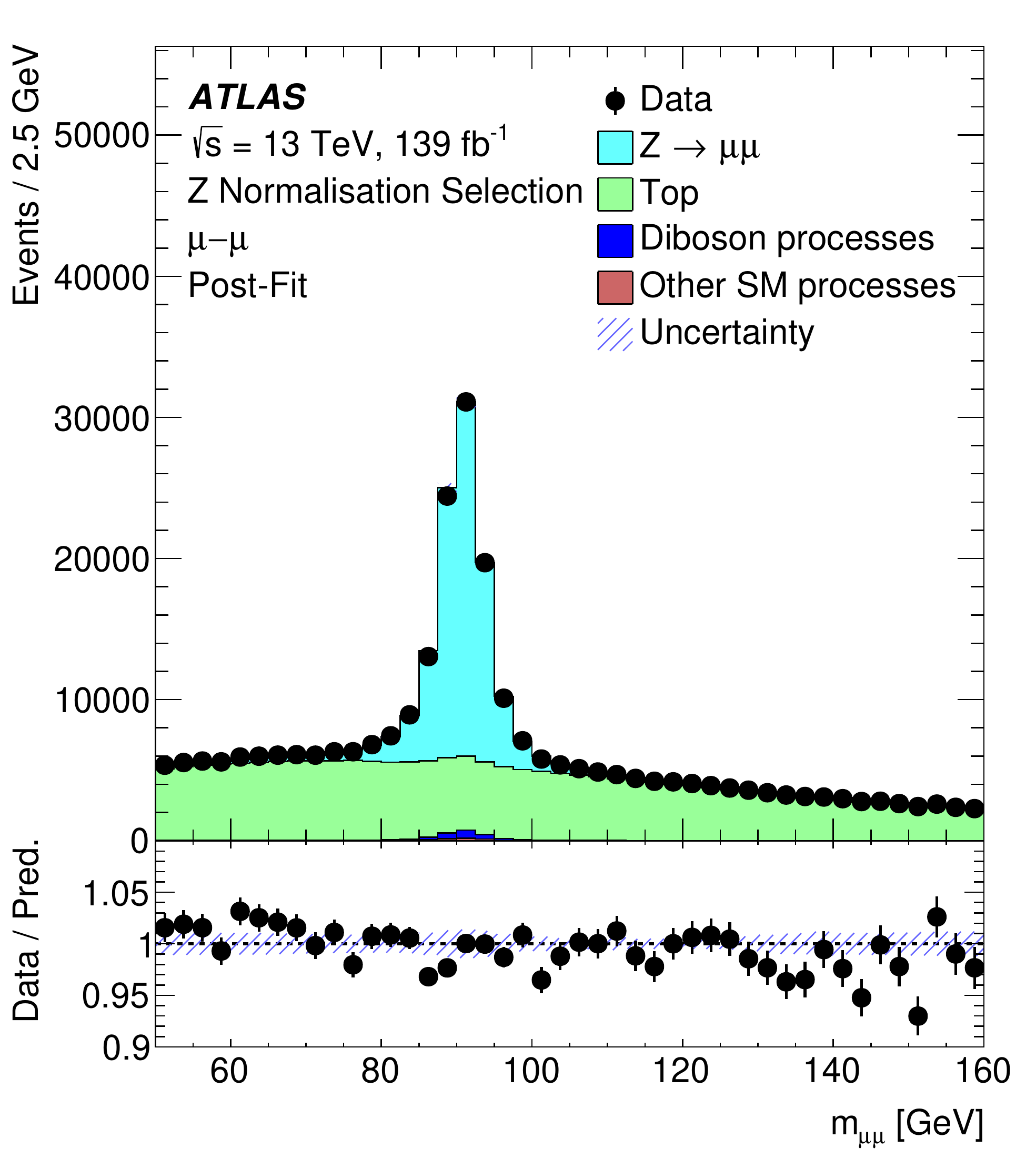}
\captionof{figure}{ \textbf{$Z\rightarrow\mu\mu$ background normalisation}
A region enriched in $Z\rightarrow\mu\mu$
events is defined by selecting events with two muons.
This control region used to
extract the $Z\rightarrow\mu\mu$~normalisation, which is applied
in the signal region. The di-muon invariant mass, $m_{\mu\mu}$,
distribution is shown in the figure. The data are shown by black
markers, the different components contributing to this region,
taken from simulation, are given by stacked histograms. The
different contributions are the primary process of interest,
$Z\rightarrow\mu\mu$, along with the main backgrounds from
events involving: top quarks, \textit{Top}; a pair of $W$ and/or $Z$
bosons, \textit{Diboson processes}; and the grouping of all
remaining Standard Model processes, \textit{Other SM processes}.
The bottom panel shows the ratio of the data to the predicted
expectation after the fit. The uncertainties on the data are the Poisson errors
due to the limited size of the data sample. Blue bands indicate
the $\pm 1\sigma$ systematic uncertainties on the prediction with the constraints from the
analysis fit applied. }
\label{fig:control.a}
\end{figure}
 
\begin{figure}
\centering
\includegraphics[width=10.2cm]{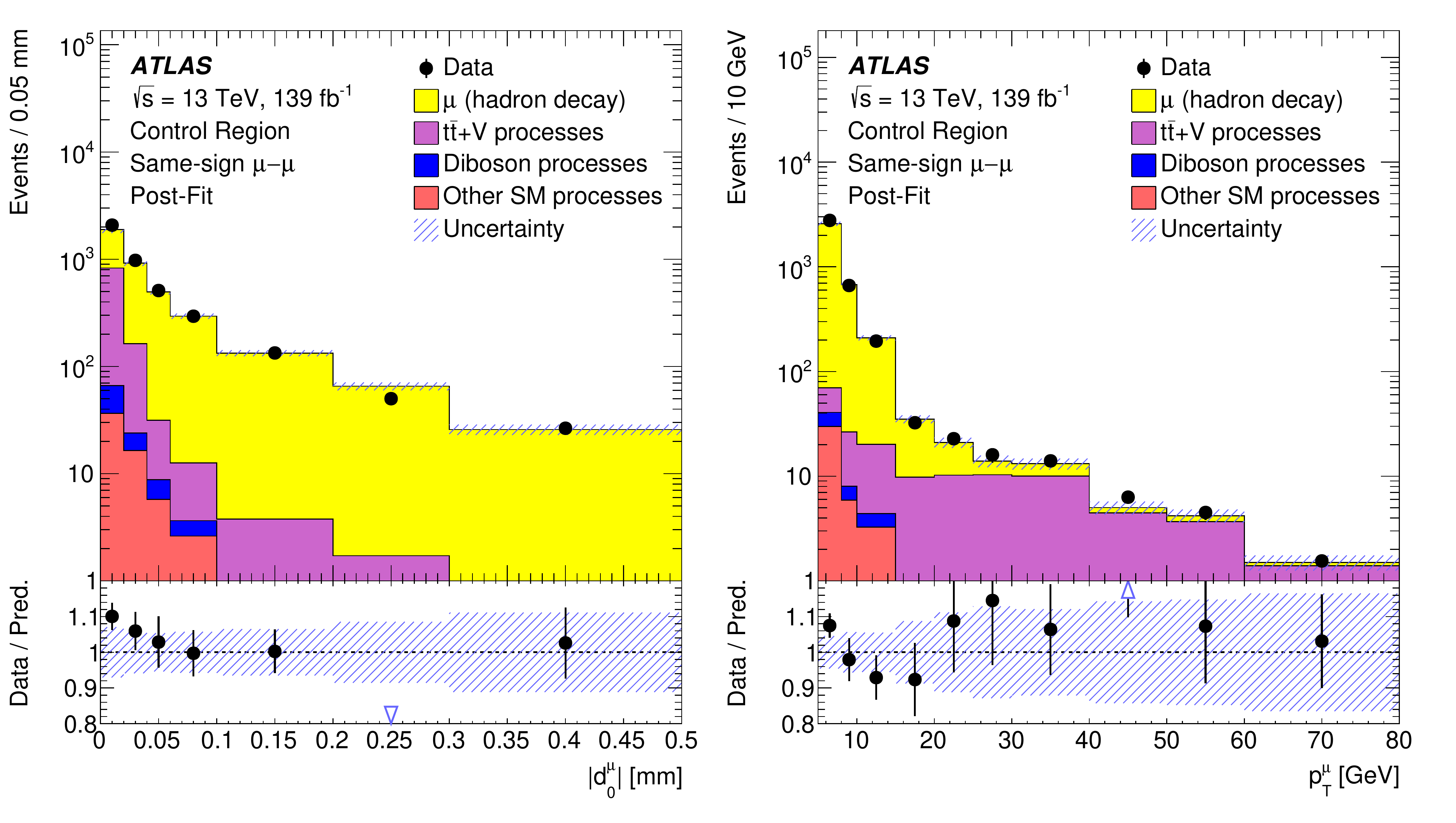}
\captionof{figure}{
\textbf{\fakemu~background normalisation}
A region enriched in \fakemu~events is defined by selecting events
with two same-sign muons. This control region used to extract
normalisation factors to correct the \fakemu~prediction to match the
data. The \textit{probe} muon tranverse impact parameter, $|d_0^\mu|$,~and
transverse momentum, $\pT^\mu$, distributions in this region are shown.  The data are
shown by black markers, the different components contributing to
this region, taken from simulation, are given by stacked
histograms. The different contributions are the primary process of
interest, \fakemu, \textit{$\mu$~(hadron~decay)}, along with the main
backgrounds from events involving: a top quark pair produced in
association with a $W$ or $Z$ boson, \textit{$\ttbar+V$ background}; a
pair of $W$ and/or $Z$ bosons, \textit{Diboson processes}; and the
grouping of all remaining Standard Model processes, \textit{Other SM
processes}.  The extracted normalisation factor is applied to the
\fakemu~prediction, along with the effect of any constraints and
pulls on the systematic uncertainties from the fit to the signal
region data.  The bottom panel shows the ratio of the data to the
predicted expectation after the fit. The uncertainties on the data are the Poisson
errors due to the limited size of the data sample. Blue bands
indicate the $\pm 1\sigma$ systematic uncertainties on the prediction with the constraints from the
analysis fit applied. The empty blue arrows in the ratio panel indicate points
where the ratio values lie outside the $y$-axis range shown.
}
\label{fig:control}
\end{figure}
 
The most important background at large values of $|d_0^\mu|$~is from
events in which the \textit{probe} muon originates from the decay of
$b$-~or $c$-hadrons, or more rarely from in-flight decays of
$\pi^\pm$~and $K^\pm$. This occurs primarily in \ttbar~events where one $W$ boson
decays leptonically and the other hadronically, referred to as
semi-leptonic \ttbar. These muons are referred to as \fakemu.  A data-driven method is used to
determine the normalisation of this background from two control
regions, one each for the $e$--$\mu$ and $\mu$--$\mu$~channels. The control regions
have the same event selection as the signal regions but the two
leptons are required to have same-sign electric charge. The
largest source of \fakemu~is from decays of $b$-hadrons, and this
contributes equally to same-sign and opposite-sign selections, and the
other significant source, $c$-hadrons, has a component in both
selections, but they are not equal. This results in a sample with
a high purity of this \fakemu~background. The extrapolation from
same-sign control region to opposite-sign signal region is estimated
from simulation. In the same-sign control region there are two
backgrounds to \fakemu~at high $\pT^\mu$: $\ttbar+V$ and \ttbar~which occurs through electron charge
misidentification in the $e$--$\mu$~channel. A
normalisation correction factor is applied to these processes
based on the number of events observed with \textit{probe} muon
$\pT^\mu>30$~GeV. This is done prior to
extracting the normalisation of the \fakemu~background. Normalisation factors to scale the simulation to
data for the \fakemu~background are found to be 1.39 (1.37) in the
$e$--$\mu$ ($\mu$--$\mu$) channels, respectively.
Figure \ref{fig:control} shows that the simulation and data are
consistent within uncertainties in the $\mu$--$\mu$~channel same-sign
control region, providing confidence that the differential
distributions of $\pT^\mu$~and $|d_0^\mu|$~are well-modelled.

% End of text imported from the .//sections/SelectCalibNorms.tex input file
% The next lines are included from the .//sections/FitResults.tex input file
\section*{Fit configuration and systematic uncertainties}\label{sec:fit}
A two dimensional profile likelihood fit \cite{Cowan:2010js}~is performed in the $|d_0^\mu|$ and $\pT^\mu$ distributions. The bin boundaries were chosen to provide the best separation between the different \promptmu, \taumu and \fakemu~processes given the available data. This resulted in three
bins in $\pT^\mu$~(boundaries of: 5, 10, 20, 250 \GeV) and eight bins
in the transverse impact parameter, $|d_0^\mu|$ (boundaries of: 0,
0.01, 0.02, 0.03, 0.04, 0.06, 0.09, 0.15, 0.5 mm), of the
\textit{probe} muon for both the $e$--$\mu$~and $\mu$--$\mu$ channels,
making 48 bins in total.
 
The ratio of the number of \taumu events to the
number \promptmu is fitted by minimizing the negative log likelihood,
to measure \Rtmu. More than 100 (nuisance) parameter values representing the
statistical and systematic uncertainties, which can be modified by the
fit, are included. The relative uncertainty of 0.23\% in the
$\tau\rightarrow\mu\nu_\mu\nu_\tau$~branching ratio is also included in the measured
value of \Rtmu and is a subdominant component of the overall
uncertainty. As both the \ttbar~and \Wt~processes contain two
$W$ bosons both are treated as signal.  Two fit parameters are allowed
to float freely: \Rtmu and $k(\ttbar)$.
The $k(\ttbar)$~parameter is a constant scaling
factor applied to the normalisation of both the \prompt and \taumu
components of the signal whilst \Rtmu only affects the
\taumu components. Both are applied across all bins and in both
channels.  \Rtmu is the parameter of interest and
it is not affected by the overall normalisation scaling factors of the
\ttbar and \Wt processes.  The fit is performed after applying the
normalisation scaling factors derived in the control regions. Other
processes are normalised to their theoretical cross-sections
taking into account the uncertainty in these predictions.

Since many systematic uncertainties are correlated between the
\promptmu and \taumu templates, they cancel out in the \Rtmu ratio,
minimising their impact on the precision of the result. These include
uncertainties related to jet reconstruction, flavour tagging and
trigger efficiencies.  The remaining dominant uncertainties are the uncertainties in
the data driven methods, the theoretical modelling uncertainties and
the reconstruction uncertainties are described in the
following, and further details are given in \textit{Methods}.

An uncertainty associated with the data-driven templates for the \absdzero
distribution of \prompt is derived to account for the differences between the $Z\rightarrow\mu\mu$~calibration region where they were derived and the signal region.

Uncertainties in the data-driven normalisation of the \fakemu~background due to
the size of the same-sign dataset, the choice of MC generators used and the uncertainty in the subtraction of the other processes in the same-sign control region are included in the fit.  The uncertainties associated with the $Z$+jets normalisation derived from data are also applied. Uncertainties on the cross-section calculations and the integrated luminosity are applied to all other backgrounds estimated from simulation, but have a minor impact on the result.

Uncertainties in the shape of the \prompt, \taumu~and \fakemu~distributions due to
the modelling of the simulated \ttbar samples are derived. The
combined yield of the \prompt and \taumu templates is allowed to float
in the fit, but changes in generator configuration choices can result
in modifications to the muon $\pT^\mu$, and subsequently the \absdzero
distribution, such that there can be relative changes in the \prompt
and \taumu template yields in each bin, leading to an uncertainty in
\Rtmu. For the \fake~background, in addition to the uncertainties in
the normalisation, there can be changes to the muon
$\pt^\mu$~modelling and the relative fractions of muons from different
sources, both of which can change the shape of the \absdzero
distribution. The most important of these variations is in the parton
shower and hadronisation model.

The muon reconstruction and isolation efficiencies are determined in
di-muon ($Z\to\mu\mu$ and $J/\psi\to\mu\mu$) data and simulation using
a \textit{tag} and \textit{probe} method~\cite{PERF-2015-10}.
Corrections are applied to the simulated samples to account for the
differences between data and simulation and the uncertainties on these
correction factors are included in the analysis.  Additionally, an
uncertainty due to the modelling of pile-up is obtained by reweighting
the simulated events~\cite{ATL-PHYS-PUB-2016-017}.

\section*{Extraction of \Rtmu}
\label{sec:results}

\begin{figure}
\begin{center}
\includegraphics[width=0.9\textwidth]{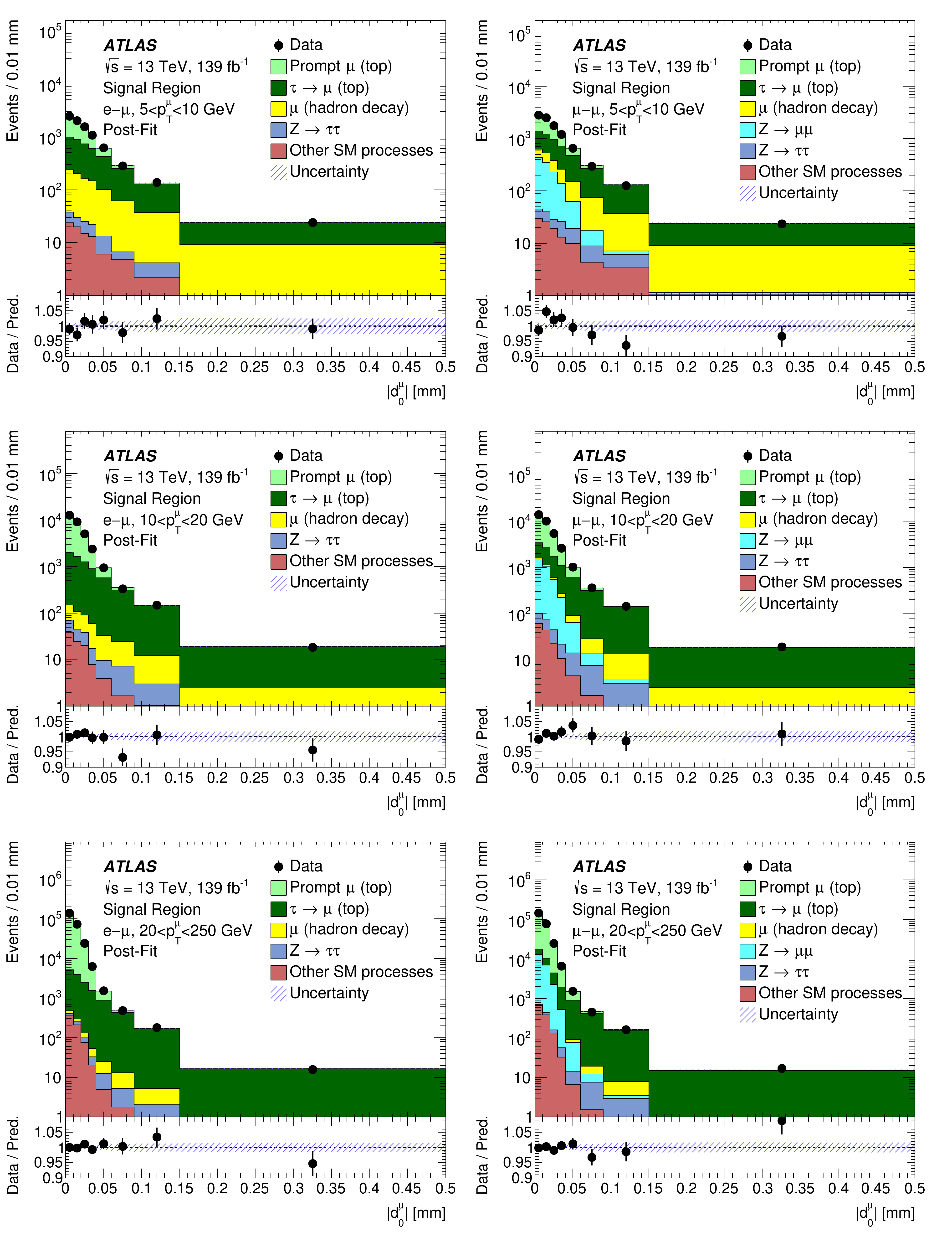}
\end{center}
\caption{\textbf{Transverse impact parameter distributions of probe muons in the signal
region} The signal region used to extract \Rtmu is enriched in
di-lepton \ttbar events.  The \absdzero~distributions for each signal region
(left: $e$--$\mu$~channel, right: $\mu$--$\mu$~channel)
and \textit{probe} muon $\pt^\mu$~bin (top: $5<\pT^\mu<10$~GeV,
middle: $10<\pT^\mu<20$~GeV, bottom: $20<\pT^\mu<250$~GeV) used in
the analysis are shown.  The data are represented by black
markers, the different components contributing to this region,
taken from simulation, are given by stacked histograms. The
different contributions are the two primary process of interest
used to extract \Rtmu, \promptmu from top quark decays, \textit{Prompt
$\mu$ (top)} and \taumu from top decays, \textit{$\tau\to\mu$
(top)}. The main backgrounds are also shown. These are: events
with a \fakemu, \textit{$\mu$~(hadron decay)}; events with a $Z$ boson
decaying to a di-muon pair, \textit{$Z\to\mu\mu$}; events with a $Z$
boson decaying to a di-$\tau$ pair, \textit{$Z\to\tau\tau$}; and the
grouping of all remaining Standard Model processes, \textit{Other SM
processes}. Distributions are shown after the fit has been
performed. The bottom panel shows the ratio of the data to the
predicted expectation after the fit. The uncertainties on the data are the
Poisson errors due to the limited size of the data sample. Blue
bands indicate the $\pm 1\sigma$ systematic uncertainties on the prediction with the constraints
from the analysis fit applied.
The empty blue arrows in the ratio panel indicate
points where the ratio values lie outside the $y$-axis range
shown. The contribution from \textit{Other SM processes} is dominated
by di-boson and $\ttbar+V$ production. The chi-square statistic
values range from 3.5 to 10.2 for 8 degrees of freedom for the
distributions.}
\label{fig:postFit}
\end{figure}

Figure~\ref{fig:postFit} shows the differential distributions of
$|d_0^\mu|$~in the six signal regions for the data and the expectation
after the fit to data.  Good agreement is observed between the
corrected simulation samples and the data. The global goodness of fit
when fitting the expectation from simulation, defined as twice the
change in negative log-likelihood relative to a fit performed assuming
the pre-fit expectation per degree of freedom, has a value of 1.11
($p$-value of 0.29).

The separation between the \prompt and \taumu processes can be seen clearly.
The \prompt processes dominate at low $|d_0^\mu|$~ while \taumu
dominates at high $|d_0^\mu|$. The \fakemu background is also
important at high $|d_0^\mu|$ but contributes most significantly at
low $\pT^\mu$.
 
The analysis was finalised prior to looking at the value of \Rtmu~in
data in order to minimise any bias. It was also checked that the result is
consistent with respect to different channels, kinematic bins,
data-taking periods and the charge of the \textit{probe} lepton.

The total systematic uncertainty is 0.011, including the uncertainty in
the $\tau\rightarrow \mu\nu_\mu\nu_\tau$~branching ratio,
and the statistical uncertainty is 0.007.
Table \ref{tab:sys} lists the different contributions of systematic
uncertainty grouped into categories. The leading contributions come
from the imperfect knowledge of the tail of the \absdzero
distribution, the parton shower and hadronisation model uncertainty,
and the muon selection uncertainties.

The measured value of \Rtmu is $$\Rtmu = 0.992 \pm 0.013~[\pm  0.007~\textrm{(stat)}~\pm~0.011~\textrm{(syst)}],$$
exceeding the precision from LEP which measured $1.070\pm0.026$. The result is shown in Figure~\ref{fig:summary} alongside the combination of LEP measurements. The present result agrees with the Standard Model expectation of equal couplings for different lepton flavours and the hypothesis of lepton-flavour universality.

\begin{table}
\centering
\caption{\textbf{Sources of uncertainty} The main
sources of uncertainty on the measured value of \Rtmu. The size of
the impact each uncertainty has on \Rtmu is assessed by fixing the
relevant fit parameters for a given uncertainty and re-fitting to
data. The reduction in the total uncertainty in this modified fit gives
the quoted impact. Different individual components used in the fit
are combined into categories.}
\begin{tabular}{|l|c|}\hline
\textbf{Source} & \textbf{Impact on \Rtmu} \\ \hline\hline
Prompt $d_0^\mu$~templates       & 0.0038 \\
\prompt and \taumus parton shower variations   & 0.0036 \\
Muon isolation efficiency   & 0.0033 \\
Muon identification and reconstruction   & 0.0030 \\
\fakemu~normalisation   & 0.0028 \\
\ttbar scale and matching variations  & 0.0027 \\
Top \pT~spectum variation   & 0.0026 \\
\fakemu~parton shower variations   & 0.0021 \\
Monte Carlo statistics   & 0.0018 \\
Pile-up   & 0.0017 \\
\taumu and \fakemu~$d_0^\mu$~shape  &  0.0017 \\
Other detector systematic uncertainties   & 0.0016 \\
$Z$+jet normalisation   & 0.0009 \\
Other sources   & 0.0004 \\ \hline
$B(\tau\rightarrow\mu\nu_\tau\nu_\mu)$ & 0.0023 \\ \hline\hline
Total systematic uncertainty & 0.0109 \\ \hline
Data statistics & 0.0072 \\ \hline\hline
\textbf{Total} & \textbf{0.013} \\ \hline
\end{tabular}
\label{tab:sys}
\end{table}

\begin{figure}[b]
\centering
\includegraphics[width=0.4\linewidth]{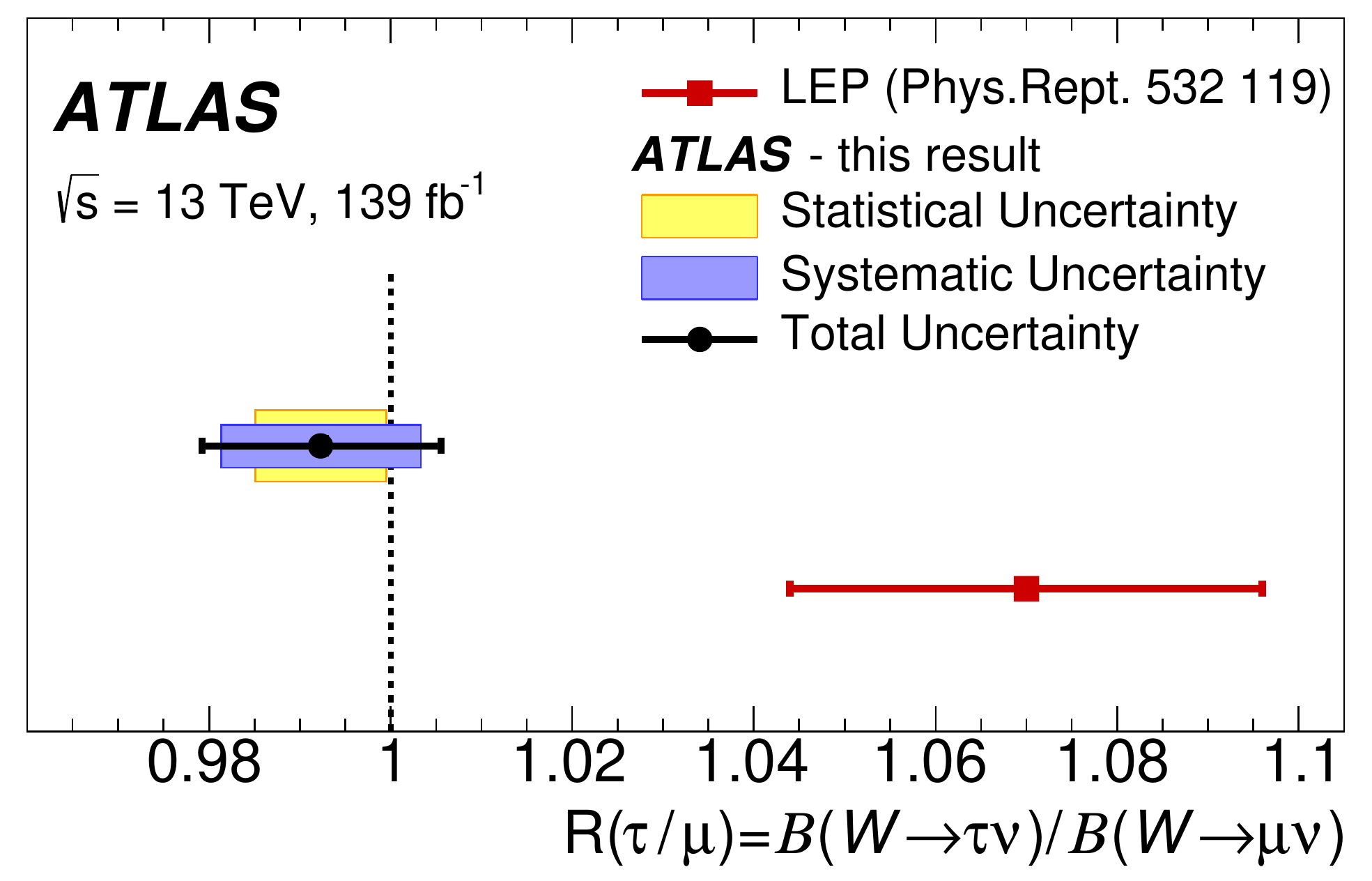}
\caption{ \textbf{Summary of ATLAS and LEP results} The new ATLAS measurement of \Rtmu and the previous LEP
result~\cite{Schael:2013ita} of the same quantity. The new measurement from ATLAS is
shown by the black circular marker and the LEP result by a red
square marker. For the ATLAS result the statistical (yellow box)
and systematic (blue box) errors are shown separately along with
the total error of the measurement (black bars). The total
uncertainty on the LEP result is indicated by the red bars. The
vertical dashed line indicates the Standard Model's prediction
lepton-flavour universality, with equal $W$ boson branching
ratios to different lepton flavours.  }
\label{fig:summary}
\end{figure}

% End of text imported from the .//sections/FitResults.tex input file
% The next lines are included from the .//sections/Conclusions.tex input file

This result surpasses the precision of the previous LEP result and resolves the
tension they observed with the SM prediction of lepton flavour
universality.  This precise measurement of $\Rtmu$ achieved so far, this is an example of the ability of the ATLAS experiment to perform high-precision measurements.
% End of text imported from the .//sections/Conclusions.tex input file
\FloatBarrier
% The next lines are included from the .//sections/Methods.tex input file
\section*{Methods}
\label{sec:methods}
 
\subsection*{ATLAS coordinate system and nomenclature}
\AtlasCoordFootnote
 
The beamline is the central ($x$--$y$) axis of the luminous region.
The transverse impact parameter, $|d_0^\mu|$, is measured in the
$x$--$y$~plane as the closest distance of approach of the track to the
beamline. It is defined relative to the beamline rather than the
primary vertex so that the resolution of $|d_0^\mu|$~is independent of
the vertex ($x$--$y$)-resolution, which depends on the physics
process.

\subsection*{Monte Carlo simulation and theoretical predictions}
\label{sec:methods-mc}

The top-pair and single-top-quark events, including $Wt$-, $t$- and $s$-channel production, were generated using the
\powhegbox~\cite{Frixione:2007nw,Nason:2004rx,Frixione:2007vw,Alioli:2010xd,Ball:2014uwa}~generator
interfaced to the
\textsc{Pythia8}~\cite{Sjostrand:2014zea,ATL-PHYS-PUB-2014-021,Ball:2012cx}
parton shower and hadronisation model, more details can be found in Ref.~\cite{ATL-PHYS-PUB-2016-020}. The decays of bottom and charm
hadrons are important for backgrounds in the analysis and were modelled using the \evtgen~\cite{EvtGen} program.
The \ttbar and single-top processes were normalised to the inclusive cross-section
calculation of the highest available precision~\cite{Beneke:2011mq,Cacciari:2011hy,Baernreuther:2012ws,Czakon:2012zr,Czakon:2012pz,Czakon:2013goa,Czakon:2011xx,Aliev:2010zk,Kant:2014oha}
and \ttbar events additionally have a differential
reweighting applied to match the next-to-next-to-leading order (NNLO) in QCD top quark \pT calculation~\cite{Czakon:2017wor}. For single top quark
production in the $Wt$-channel, the diagram removal
scheme~\cite{Frixione:2008yi} was used to remove overlap with
\ttbar\ production. $\ttbar+W$ and $\ttbar+Z$ events were simulated with \sherpa and \mgamc~\cite{Alwall:2014hca} interfaced with \textsc{Pythia8}, respectively.
The background from $V (=W,Z)+$jets and $VV$+jets events was simulated with the \sherpa~\cite{Gleisberg:2008ta,Bothmann:2019yzt,Gleisberg:2008fv,Cascioli:2011va,Denner:2016kdg,Schumann:2007mg,Hoeche:2011fd,Hoeche:2012yf,Catani:2001cc,Hoeche:2009rj,Ball:2014uwa} generator, more details can be found in Refs.~\cite{ATL-PHYS-PUB-2017-006,ATL-PHYS-PUB-2017-005}. All processes were normalised to their highest order available cross sections~\cite{Anastasiou:2003ds,deFlorian:2016spz}.

Simulated inelastic $pp$ collisions~\cite{ATL-PHYS-PUB-2016-017} were overlaid on events in all samples to model the observed data distribution of pile-up from addition collisions in the same and neighbouring bunch crossings.

\subsection*{Object identification in ATLAS}
\label{sec:methods-objects}

Muons are reconstructed using combined fits of inner detector \cite{PERF-2015-08,ATL-PHYS-PUB-2015-051}~and muon spectrometer tracks~\cite{PERF-2015-10}.
They are required to satisfy the `medium' identification critera of Ref.~\cite{PERF-2015-10}. They are also required to be strictly isolated from other activity by requiring that the sum of the \pT of other tracks within a surrounding cone of size $\Delta R=0.3$ and the sum of \pT calculated from calorimeter energy deposits within a cone of size $\Delta R=0.2$ around the muons are below certain thresholds. \textit{Tag} muons are required to have $\pT^\mu > 27.3$~GeV to pass the trigger thresholds, while \textit{probe} muons are required to have $\pT^\mu > 5$~GeV. Both the \textit{tag} and \textit{probe} muons are required to have $|\eta|<2.5$~and to originate close to the primary vertex with a distance of closest approach in the $r$--$z$~plane of less than 0.3 mm and a transverse impact parameter relative to the beamline, $|d_0^\mu|$, of less than 0.5 mm. The primary vertex is defined as the vertex with the highest $\Sigma\pT^2$~of the tracks associated with it. Additional criteria are applied to test the compatibility of the momenta measured separately in the inner detector and the muon spectrometer, in order to remove reconstructed muons which result from in-flight decays of charged kaons and pions.
 
Electrons are reconstructed from inner detector tracks matched to clusters of calorimeter-cell energy clusters~\cite{EGAM-2018-01}. They are required to satisfy the `tight' identification criteria of Ref.~~\cite{EGAM-2018-01} and the same strict isolation criteria as applied to muons. \textit{Tag} electrons are required to have $\pT^e>27$~GeV, to pass the trigger requirements and satisfy $|\eta|<2.47$, excluding the transition region between the barrel and end-cap calorimeter, $1.37<|\eta|<1.52$. They must also satisfy the same criteria as for muons for their distance of closest approach to the primary vertex in the transverse and $r$--$z$~plane.
 
Hadronic jets are built from the energy in calorimeter cells at the electromagnetic energy scale~\cite{PERF-2014-07}, using the anti-$k_t$~algorithm~\cite{Cacciari:2008gp} with a radius parameter of 0.4. They are then calibrated to the energy scale of jets created from stable generator-level particles excluding muons and neutrinos using the same jet algorithm~\cite{PERF-2016-04}. 
For jets with $25<\pT<60$~GeV~and $|\eta|<2.4$, pile-up suppression requirements in the form of a jet vertex tagger \cite{PERF-2014-03} are applied. Only jets with $\pT>25$~GeV and $|\eta|<2.5$~are considered in the analysis. To classify jets as containing a $b$-hadron, the 70\% efficiency working point of the MV2c10 $b$-tagging algorithm \cite{PERF-2016-05,FTAG-2018-01} is used.
 
An overlap removal procedure as described in Ref.~\cite{TOPQ-2018-01} is applied to resolve the ambiguity if lepton signals in the calorimeter are also reconstructed as hadronic jets.

\subsection*{Systematic uncertainties}
 
Several of the most important systematic uncertainties in the
measurement of \Rtmu merit a more detailed discussion, which is
provided here.
 
Due to differences in the hadronic environment around the lepton
between $Z$ and \ttbar final states and the coarse binning in
$\pT^\mu$~and $|\eta|$, which may not be able to encapsulate the full
shape information, small biases can exist in the data-driven \prompt template
distributions. The size of such a possible bias is estimated from the
full difference between \prompt \absdzero templates from $Z$ and
\ttbar in simulation. This uncertainty is split into two components
corresponding to the tail, $\absdzero\gtrsim0.05$~mm, and core,
$\absdzero\lesssim0.05$~mm, to prevent the data from constraining the
uncertainty by using the full \absdzero distribution.
 
The breakdown of uncertainties contributing to the \fakemu~background control region normalisation
in the $e$--$\mu$ ($\mu$--$\mu$) channels is: 4\% (4\%) due to the
size of the same-sign dataset; 8\% (3\%) due to the choice of MC
generators used; and 1.0\% (1.3\%) due the uncertainty in the
subtraction of the other processes in the same-sign control region.

The uncertainties due to the choice of MC event generator for the
\prompt, \taumu and \fakemu~processes are estimated by varying
different components of the modelling in a factorised way. The following variations are considered targetting different sources of uncertainty:
\begin{itemize}
\item \textbf{Inital- and final-state radiation:} A14 eigen-tune variations~\cite{ATL-PHYS-PUB-2014-021} of the strong coupling ($\alpha_{\textrm{s}}$),
\item \textbf{Missing higher-order QCD corrections:} Factorisation and renormalisation scales simultaneously varied up and down by a factor of two,
\item \textbf{Resummation scale uncertainty:} \powheg~$h_{\textrm{damp}}$~parameter varied from 1.5 to 3~\mtop (symmetrised),
\item \textbf{Parton shower and hadronisation model:} \herwig~7~\cite{Bahr:2008pv,Bellm:2015jjp}, H7UE tune~\cite{Bellm:2015jjp,Harland-Lang:2014zoa} (symmetrised),
\item \textbf{Higher-order correction to top \pT~spectrum:} Not applying the NNLO top \pT~reweighting~\cite{Czakon:2017wor} (symmetrised).
\end{itemize}
In the cases where only a single alternative is
given, the uncertainty is taken to be the deviation from the nominal
result and then symmetrised, as indicated above.
The effects on \prompt and \taumu
(collectively referred to as `signal') are treated as correlated and the effects on
\fakemu~are treated separately.
The parton shower and hadronisation uncertainty is
separated into four nuisance parameters: one each corresponding to low and
middle probe-muon $\pt^\mu$~bins used in the fit, and two corresponding to
the high $\pt^\mu$~bin where the uncertainty is further separated into
components related to normalisation and shape differences.
% End of text imported from the .//sections/Methods.tex input file
 
\section*{Acknowledgements}
 
% The next lines are included from the .//acknowledgements/Acknowledgements.tex input file

We thank CERN for the very successful operation of the LHC, as well as the
support staff from our institutions without whom ATLAS could not be
operated efficiently.
 
We acknowledge the support of
ANPCyT, Argentina;
YerPhI, Armenia;
ARC, Australia;
BMWFW and FWF, Austria;
ANAS, Azerbaijan;
SSTC, Belarus;
CNPq and FAPESP, Brazil;
NSERC, NRC and CFI, Canada;
CERN;
ANID, Chile;
CAS, MOST and NSFC, China;
Minciencias, Colombia;
MSMT CR, MPO CR and VSC CR, Czech Republic;
DNRF and DNSRC, Denmark;
IN2P3-CNRS and CEA-DRF/IRFU, France;
SRNSFG, Georgia;
BMBF, HGF and MPG, Germany;
GSRI, Greece;
RGC and Hong Kong SAR, China;
ISF and Benoziyo Center, Israel;
INFN, Italy;
MEXT and JSPS, Japan;
CNRST, Morocco;
NWO, Netherlands;
RCN, Norway;
MNiSW and NCN, Poland;
FCT, Portugal;
MNE/IFA, Romania;
JINR;
MES of Russia and NRC KI, Russian Federation;
MESTD, Serbia;
MSSR, Slovakia;
ARRS and MIZ\v{S}, Slovenia;
DSI/NRF, South Africa;
MICINN, Spain;
SRC and Wallenberg Foundation, Sweden;
SERI, SNSF and Cantons of Bern and Geneva, Switzerland;
MOST, Taiwan;
TAEK, Turkey;
STFC, United Kingdom;
DOE and NSF, United States of America.
In addition, individual groups and members have received support from
BCKDF, CANARIE, Compute Canada and CRC, Canada;
COST, ERC, ERDF, Horizon 2020 and Marie Sk{\l}odowska-Curie Actions, European Union;
Investissements d'Avenir Labex, Investissements d'Avenir Idex and ANR, France;
DFG and AvH Foundation, Germany;
Herakleitos, Thales and Aristeia programmes co-financed by EU-ESF and the Greek NSRF, Greece;
BSF-NSF and GIF, Israel;
Norwegian Financial Mechanism 2014-2021, Norway;
La Caixa Banking Foundation, CERCA Programme Generalitat de Catalunya and PROMETEO and GenT Programmes Generalitat Valenciana, Spain;
G\"{o}ran Gustafssons Stiftelse, Sweden;
The Royal Society and Leverhulme Trust, United Kingdom.
 
The crucial computing support from all WLCG partners is acknowledged gratefully, in particular from CERN, the ATLAS Tier-1 facilities at TRIUMF (Canada), NDGF (Denmark, Norway, Sweden), CC-IN2P3 (France), KIT/GridKA (Germany), INFN-CNAF (Italy), NL-T1 (Netherlands), PIC (Spain), ASGC (Taiwan), RAL (UK) and BNL (USA), the Tier-2 facilities worldwide and large non-WLCG resource providers. Major contributors of computing resources are listed in Ref.~\cite{ATL-SOFT-PUB-2021-003}.
 
% End of text imported from the .//acknowledgements/Acknowledgements.tex input file
 
\clearpage
\section*{Author contributions}
All authors have contributed to the publication, being variously involved in the design and the construction of the detectors, in writing software, calibrating subsystems, operating the detectors and acquiring data, and finally analysing the processed data. The ATLAS Collaboration members discussed and approved the scientific results. The manuscript was prepared by a subgroup of authors appointed by the collaboration and subject to an internal collaboration-wide review process. All authors reviewed and approved the final version of the manuscript.
 
\section*{Competing interests}
 
The authors declare no competing interests.
 
\section*{Data availability}
 
The experimental data that support the findings of this study are available in
HEPData with the identifier \url{https://www.hepdata.net/record/100232}.
The ATLAS software is available at the following link: \url{https://gitlab.cern.ch/atlas/athena}.

\clearpage

\printbibliography

\clearpage
% ATLAS Collaboration author list
% Reference date of TOPQ-2018-29 is 2020-05-08
% Author list last updated on date 20-JAN-21
% Data extracted on 20-Jan-2021 for paper reference TOPQ-2018-29
% at 11:49am
 
\begin{flushleft}
\hypersetup{urlcolor=black}
{\Large The ATLAS Collaboration}

\bigskip

\AtlasOrcid[0000-0002-6665-4934]{G.~Aad}$^\textrm{\scriptsize 102}$,    
\AtlasOrcid[0000-0002-5888-2734]{B.~Abbott}$^\textrm{\scriptsize 128}$,    
\AtlasOrcid[0000-0002-7248-3203]{D.C.~Abbott}$^\textrm{\scriptsize 103}$,    
\AtlasOrcid[0000-0002-2788-3822]{A.~Abed~Abud}$^\textrm{\scriptsize 36}$,    
\AtlasOrcid[0000-0002-1002-1652]{K.~Abeling}$^\textrm{\scriptsize 53}$,    
\AtlasOrcid[0000-0002-2987-4006]{D.K.~Abhayasinghe}$^\textrm{\scriptsize 94}$,    
\AtlasOrcid[0000-0002-8496-9294]{S.H.~Abidi}$^\textrm{\scriptsize 167}$,    
\AtlasOrcid[0000-0002-8279-9324]{O.S.~AbouZeid}$^\textrm{\scriptsize 40}$,    
\AtlasOrcid{N.L.~Abraham}$^\textrm{\scriptsize 156}$,    
\AtlasOrcid[0000-0001-5329-6640]{H.~Abramowicz}$^\textrm{\scriptsize 161}$
% ...
\bigskip
\\

% ...
$^{36}$CERN, Geneva; Switzerland.\\
% ...
$^{40}$Niels Bohr Institute, University of Copenhagen, Copenhagen; Denmark.\\
% ...
$^{53}$II. Physikalisches Institut, Georg-August-Universit\"{a}t G\"ottingen, G\"ottingen; Germany.\\
% ...
$^{94}$Department of Physics, Royal Holloway University of London, Egham; United Kingdom.\\
% ...
$^{102}$CPPM, Aix-Marseille Universit\'e, CNRS/IN2P3, Marseille; France.\\
$^{103}$Department of Physics, University of Massachusetts, Amherst MA; United States of America.\\
% ...
$^{128}$Homer L. Dodge Department of Physics and Astronomy, University of Oklahoma, Norman OK; United States of America.\\
% ...
$^{156}$Department of Physics and Astronomy, University of Sussex, Brighton; United Kingdom.\\
% ...
$^{161}$Raymond and Beverly Sackler School of Physics and Astronomy, Tel Aviv University, Tel Aviv; Israel.\\
% ...
$^{167}$Department of Physics, University of Toronto, Toronto ON; Canada.\\
% ...

\end{flushleft}

% Created with Glance <Atlas.Glance@cern.ch>

\end{document}